\documentclass{mn2e}

\usepackage[dvips]{graphicx}
\usepackage{amssymb,amsbsy,txfonts}

\newcommand{\msun}{{\rm M}_{\sun}}

\newcommand{\apj}{{ApJ}}
\newcommand{\mnras}{{MNRAS}}
\newcommand{\aap}{{A\&A}}
\renewcommand\vec[1]{\ensuremath\boldsymbol{#1}}
\def\ee{\end{equation}}
\def\be{\begin{equation}}
\newcommand{\ledd}{L_{{\rm E}}}
\newcommand{\medd}{{\dot{M}_{\rm E}}}

\title[Global Compton cooling of hot accretion flows]{Monte Carlo simulations of global Compton cooling in inner regions of hot accretion flows}

\author[F. G.\ Xie, A. Nied\'zwiecki, A. A. Zdziarski and F. Yuan]
{Fu-Guo Xie,$^{1,4}$\thanks{E-mail: fgxie@shao.ac.cn (FGX), niedzwiecki@uni.lodz.pl (AN), aaz@camk.edu.pl (AAZ), fyuan@shao.ac.cn (FY)} Andrzej Nied\'zwiecki,$^2$\footnotemark[1] Andrzej A.~Zdziarski$^{3}$\footnotemark[1] and Feng Yuan$^1$\footnotemark[1]\\
$^1$Key Laboratory for Research in Galaxies and Cosmology, Shanghai Astronomical Observatory, Chinese Academy of Sciences,\\
80 Nandan Road, Shanghai 200030, China\\
$^2$Department of Astrophysics, University of \L \'od\'z, Pomorska 149/153, 90-236 \L \'od\'z, Poland\\
$^{3}$ Centrum Astronomiczne im.\ M. Kopernika, Bartycka 18, 00-716 Warszawa, Poland\\
$^4$Kavli Institute for Astronomy and Astrophysics, Peking University, Beijing 100871, China\\
}

\pagerange{\pageref{firstpage}--\pageref{lastpage}}
\pubyear{2009}

\begin{document}

\maketitle

\label{firstpage}

\begin{abstract}
Hot accretion flows such as advection-dominated accretion flows are generally optically thin in the radial direction. Thus photons generated at some radii can cool or heat electrons at other radii via Compton scattering. Such global Compton scattering has previously been shown to be important for the dynamics of accretion flows. Here, we extend previous treatments of this problem by using accurate global general relativistic Monte Carlo simulations. We focus on an inner region of the accretion flow ($R\leq 600R_{\rm g}$), for which we obtain a global self-consistent solution. As compared to the initial, not self-consistent solution, the final solution has both the cooling rate and the electron temperature significantly reduced at radii $\ga 10$ gravitational radii. On the other hand, the radiation spectrum of the self-consistent solution has the shape similar to that of the initial iteration, except for the high-energy cut-off being at an energy lower by a factor of $\sim\! 2$ and the bolometric luminosity decreased by a factor of $\sim\! 2$. We also compare the global Compton scattering model with local models in spherical and slab geometry. We find that the slab model approximates the global model significantly better than the spherical one. Still, neither local model gives a good approximation to the radial profile of the cooling rate, and the differences can be up to two orders of magnitude. The local slab model underestimates the cooling rate at outer regions whereas it overestimates that rate at inner regions. We compare our modelling results to observed hard-state spectra of black-hole binaries and find an overall good agreement provided any disc outflow is weak. We find that general-relativistic effects in flows which dynamics is modified by global Comptonization is crucial in approaching this agreement.
\end{abstract}

\section{Introduction}
\label{intro}

The lowest-order interaction between photons and electrons, namely
Compton scattering, results in momentum and energy exchange. It is a
common and important process in astrophysics. The momentum exchange
regulates accreting gases to accrete only below a critical value,
corresponding to the Eddington limit on the luminosity, $\ledd$ (for
spherical accretion). The importance of energy exchange manifests itself in
two aspects. For photons, Compton up-scattering by energetic
electrons is the main mechanism to produce X-ray emission in various
astrophysical systems, and Compton down-scattering by low-energy
electrons often leads to modification of X-ray spectra. For
electrons, they either gain or lose energy, depending on the average
photon energy. This process is important, in particular, in hot
accretion flows.

Hot accretion flows such as advection-dominated accretion flows
(ADAF, e.g., Narayan \& Yi 1994; Abramowicz et al.\ 1995) are
optically thin in both vertical and radial directions. Thus, a
photon can travel a long distance before being absorbed or scattered
(e.g., Narayan \& Yi 1994; Narayan, Mahadevan \& Quataert 1998).
Then, photons produced in one location can either cool or heat the
flow in another location, and Compton effect is not local but {\it
global}. The extension of ADAF to higher accretion rates (but still
below the Eddington limit), namely luminous hot accretion flows (LHAF, Yuan
2001 hereafter Y01; Yuan 2003), should be even more affected by the global Compton scattering, since its optical depths are higher and the radiation fields are stronger than those of an ADAF.

Global Compton cooling effects due to external photon sources
like the surface of a neutron star have been investigated in the past (e.g., Narayan \& Yi 1995). In this study, we concentrate on pure hot
accretion flows (ADAF or LHAF) where all the seed photons for
Compton scattering processes are generated within the hot plasma. We note that if the geometry of an accretion flow is an inner hot flow plus an outer truncated thin disc (e.g., Esin, McClintock \& Narayan 1997), cooling by the outer thin disc (Shakura \& Sunyaev 1973) should also be included; however, we neglect it here for simplicity.

While Comptonization in the vertical direction is included in the
standard ADAF/LHAF models by a local one-zone treatment (e.g.,
Narayan \& Yi 1995; Manmoto, Mineshige \& Kusunose 1997), any non-local radiative transfer in the radial direction, especially Comptonization, has been
neglected in almost all of the previous works in this field. Global
Comptonization in hot accretion flows has been considered, as far as
we know, only by Esin (1997), Kurpiewski \& Jaroszy\'nski (1999),
Park \& Ostriker (1999, 2001, 2007), and recently by Yuan, Xie \&
Ostriker (2009, hereafter YXO09). We note that only YXO09 and
Kurpiewski \& Jaroszy\'nski (1999) attempted to find a global
solution to this problem rather than to use a self-similar
approach, which fails at the inner regions, where most of the
radiation is generated. YXO09 used an iteration method to find, for
the first time, the radiative cooling rate mutually consistent
between the dynamics and radiation of the flow. They have found that
Compton heating and cooling dominates outside and inside,
respectively, of $\sim\! 10^4 R_{\rm g}$, where $R_{\rm g} = G
M_{\rm BH}/c^2$ is the gravitational radius and $M_{\rm BH}$ is the black-hole mass. They have also found that the relative effect of global Comptonization increases with the increasing accretion rate.

We note, however, that YXO09 have concentrated on the Compton
heating of the outer flow. Also, the only photon
production rate they used was from the intrinsic emission
(synchrotron and bremsstrahlung) and their local Comptonization
at a given radius. Thus, although they did calculate the effect
of the Compton energy exchange (using the exact relativistic
treatment of Guilbert 1986) on the electron temperature in a
given radius including photons produced at all radii, they did
not take into account the photons generated after the non-local
scattering. Thus, non-locally scattered photons were assumed to
leave the flow. Since outer parts of hot flows are very
optically thin, this is a reasonable assumption for calculating
their Compton heating.

On the other hand, here we consider Compton cooling of an inner part of a hot flow, where the Thomson optical depth becomes comparable to unity at moderate
accretion rates. Thus, we cannot use the method of YXO09. Instead,
we use a Monte Carlo (hereafter MC) method for Comptonization in the hot flow. Our method is directly based on that of Nied\'zwiecki (2005) and
Nied\'zwiecki \& Zdziarski (2006, hereafter NZ06), which is a
generalization of the method of Pozdnyakov, Sobol' \& Sunyaev (1983)
and G\'orecki \& Wilczewski (1984) to include effects of special and
general relativity (hereafter GR) and the bulk motion of the flow.
Comptonization by bulk motion (considering inflow only) was earlier
studied by, e.g., Titarchuk, Mastichiadis \& Kylafis (1997) and
Laurent \& Titarchuk (1999; see also NZ06 for a critical discussion). We assume that electrons have a relativistic Maxwellian distribution, without any non-thermal contribution. In our dynamical treatment, we use a one-dimensional model, which assumes the flow to be both azimuthally and tangentially uniform, with the scale height given by hydrostatic equilibrium.

MC simulations for Compton scattering process in a hot
accretion flow in the Kerr metric was first done by
Kurpiewski \& Jaroszy\'nski (1999). However, their dynamical
model was based on the fully advection-dominated flow (i.e.,
with negligible radiative cooling), which is clearly not the
case for flows with higher accretion rates, which we investigate
here.

We describe our treatment of the problem in Section \ref{model}. We present a comparison of various Comptonization models in Section \ref{compmodels}. In Section \ref{adaf}, we discuss the effects of outflows and viscous electron heating on the accretion solutions. The self-consistent solution for our chosen accretion rate is discussed in Section \ref{results}. Our results are compared with observations in Section \ref{comparison}, and conclusions are given in Section \ref{conclusions}.

\section{The model}
\label{model}

Here, we state the problem we are solving, and present the dynamical equations and the Compton scattering method we use. We focus on a hot accretion flow around a stellar black hole with $M_{\rm BH} = 10 \msun$. The outer boundary is set at $R_{\rm out} = 600 R_{\rm g}$ and the accretion rate there is $\dot M_0$.

In order to study the effect of global Comptonization, we need
to know the structure of the flow, i.e., the density,
temperature and velocity as functions of the radius. We iterate
between the dynamical solutions and the MC Comptonization
results to obtain the self-consistent Compton cooling rate. We
start from solving the dynamical structure without considering
the global Comptonization, which is customary in most of the
previous work on hot accretion flows (e.g., Narayan \& Yi 1995;
Manmoto et al. 1997; Quataert \& Narayan 1999; Yuan, Quataert \&
Narayan 2003). Then we calculate the radially-dependent Compton
cooling by the MC simulations. We then use this rate to
calculate again the dynamical structure. We repeat the above
procedure until it converges.

\subsection{Dynamical equations of hot accretion flows}
\label{dynsec}

The scale height, $H$, of the hot accretion flow in our solution depends on the radius. It is given by hydrostatic equilibrium,
\begin{equation}
H=\left(p\over \rho\right)^{1/2}{1\over \Omega_{\rm K}},
\label{eq:hydro}
\end{equation}
where $p$ is the total pressure, which includes contributions from
electrons, ions and magnetic field (see below), and $\rho$ is the
mass density, where we assume cosmic abundances with the H mass fraction of $X=0.75$. We assume that the density is tangentially uniform from the disc mid-plane up to $\pm H$. We adopt the pseudo-Newtonian potential of Paczy\'nski \& Wiita (1980), in which $\Omega_{\rm K}$ (the Keplerian angular velocity) is given by,
\begin{equation}
\Omega_{\rm K} = \left[G M_{\rm BH}\over R(R-2 R_{\rm
g})^2\right]^{1/2}. \label{eq:Omega_K}
\end{equation}

Because of the effect of outflow/convection, the mass inflow
rate is function of radius and we assume it to be given by
\be
\dot M_{\rm inflow} = -4\upi R H\rho v = \dot
M_0\left(\frac{R}{R_{\rm out}}\right)^s, \label{eq:outflow}
\ee
where $v$ ($<0$ for the inflow) is the radial velocity. Hereafter, we adopt the unit of $\medd=\ledd/c^2$ for $\dot M_0$, where $\ledd\equiv 4\upi GM_{\rm BH} m_{\rm p} c/\sigma_{\rm T}$ is the Eddington luminosity for pure H, $m_{\rm p}$ is the proton mass and $\sigma_{\rm T}$ is the Thomson cross section. The value of $s$ is well constrained in the case of the supermassive black hole in our Galactic Centre (Yuan, Quataert \& Narayan 2003), and we set $s = 0.3$ following that work. We discuss the validity and consequences of this assumption in Sections \ref{adaf} and \ref{comparison} below.

The energy equation for electrons is,
\be \rho v \left(\frac{{\rm
d} \varepsilon_{\rm e}}{{\rm d}R}- {p_{\rm e} \over \rho^2}
\frac{{\rm d} \rho}{{\rm d}R}\right) =\delta q_{\rm vis} + q_{\rm
ie}-q_{\rm syn+br} -q_{\rm comp}, \label{eq:energy}
\ee
where $\varepsilon_{\rm e}$ is the specific internal energy of electrons,
$p_{\rm e}$ is the electron pressure, and $q_{\rm ie}$ is the
electron heating rate per unit volume by ions via Coulomb
collisions. The electrons can also be directly heated at the rate
$\delta q_{\rm vis}$, where $q_{\rm vis}$ is the viscous heating
rate per unit volume. We assume $\delta=0.5$, again following the
detailed study of the Galactic Centre accretion flow of Yuan et al.\ (2003). The electrons are cooled by synchrotron
and bremsstrahlung, $q_{\rm syn+br}$, and by the Compton scattering
process, $q_{\rm comp}$. The latter is the net Compton cooling rate, which is, by definition,
\begin{equation}
q_{\rm comp} = n_{\rm e} \int \sigma(E, T_{\rm e}) {J_E\over E} \left(\langle
E_1\rangle-E\right) {\rm d}E.
\label{eq:Compton}
\end{equation}
Here, $J_E$ is the mean intensity per unit photon energy, $E$, in
the local rest frame at $R$, $\sigma(E, T_{\rm e})$ is the
Compton cross section averaged over the electron Maxwellian
distribution at the temperature, $T_{\rm e}$, and $\langle E_1\rangle$ is the average photon energy (in the same frame) after scattering. However, since we
use MC simulations (see Section \ref{montecarlo}) to find our
self-consistent solution, we calculate the (equivalent)
Compton energy exchange rate directly from the simulations
instead of using equation (\ref{eq:Compton}). This rate differs
from the local Compton cooling rate (commonly used in other
studies of hot accretion flows) in that $J_E$ includes now
photons either produced or scattered at all radii that reach a
given point at $R$. It can also become negative, i.e.,
corresponding to the radiative heating by photons, especially in
outer regions of the flow ($\ga 10^4 R_{\rm g}$; e.g., YXO09).
However, since we concentrate here on inner flows, we find
$q_{\rm comp} > 0$ in all cases considered here.

We assume the ratio of the gas pressure (electron and ion) to the
magnetic pressure, $\beta_{\rm B}=9$, i.e., the magnetic pressure of a 1/10th of the total pressure. This parameter determines the strength of the magnetic field in the accretion flow, and the synchrotron emission can then be determined. For it, we follow the method of Narayan \& Yi (1995) and Manmoto et al.\ (1997).

For completeness, we list below the other dynamical equations,
i.e., equations for the radial and angular momenta and for the
ion energy (see, e.g., Yuan et al.\ 2003; Manmoto et al.\ 1997),
\begin{eqnarray}
\lefteqn{ v\frac{{\rm d}v}{{\rm d}R} =  -\Omega_{\rm K}^2 R + \Omega^2 R
-{1\over \rho} \frac {{\rm d} p}{{\rm d}R},}\\
\lefteqn{v (\Omega R^2-j)  =  - {\alpha R p/ \rho},}\\
\lefteqn{\rho v \left(\frac{{\rm d} \varepsilon_{\rm i}}{{\rm d}R}-
{p_{\rm i} \over \rho^2} \frac{{\rm d} \rho}{{\rm d}R} \right)
= (1-\delta)q_{\rm vis} - q_{\rm ie},} \label{eq:adaf}
\end{eqnarray}
where $\Omega$ is the angular velocity of the flow, $j$ is the
specific angular momentum of the accretion flow when it crosses the horizon, and $\alpha$ is the viscous parameter, which we set here as $\alpha=0.3$.

We note that except for using the pseudo-Newtonian potential,
our dynamical equations are non-relativistic (NR). The radial velocity
then un-physically exceeds the speed of light in the vicinity of
the black hole. We have tried two kinds of corrections for this.
First, in modelling the Galactic Centre, Yuan, Shen \& Huang
(2006) divide the velocity by factor $0.93 \exp(4.26 R_{\rm
g}/R)$, which comes from fitting to corresponding relativistic
results. Second, we note that a first-order dynamical
relativistic correction is to replace the NR momentum, $mv$,
by $\gamma mv$, where $\gamma$ is the Lorentz factor of the
flow. We then approximate the actual velocity using the
assumption that the velocities in the NR calculations are the
actual ones multiplied by factor $\gamma$. We have compared
these two methods and found no significant differences. Here, we
use the latter treatment.

\subsection{Monte Carlo method for Compton scattering}
\label{montecarlo}

The MC method used here is fully GR in the Kerr metric, and it is described in Nied\'zwiecki (2005) and NZ06. However, in the present paper we consider only a non-rotating black hole. We thus use the following dimensionless
parameters,
\begin{equation}
r = {R \over R_{\rm g}}, \quad \vec{\lambda} = { \vec{L} c \over E_{\rm inf}
R_{\rm g}}, \quad\lambda_{\phi} = {L_{\phi} c \over E_{\rm
inf} R_{\rm g}},\quad \beta^r = {v^r \over c},\quad \beta^{\phi} \equiv
{v^\phi \over c},
\end{equation}
where ${\vec L} = {\vec R} \times {\vec p}$ is the vector of the photon angular momentum, ${\vec p} = (p_R,p_{\theta},p_{\phi})$ is the photon momentum vector, ${\vec R} = (R,0,0)$, $L_{\phi}$ is the component of $\vec{L}$
parallel to the symmetry axis, i.e., $L_{\phi}=L \cos \xi$, and $\xi$ is the (dihedral) angle between the plane of photon motion and the equatorial plane. Then, $E_{\rm inf}$ is the photon energy at infinity, $v^r$ and $v^{\phi}$ are the radial and azimuthal velocities of the flow, respectively, with respect to local static observers (i.e., observers with $r$, $\theta$ and $\phi$
= constant). We use the relativity-corrected (as described above) values of $v$ and $R\Omega$ as $v^r$ and $v^\phi$, respectively.

The tangential motion of the accretion flow is neglected,
$v^\theta=0$. The tangential (or vertical) position of a
photon within the flow is given by $\theta$. The energy
generation and exchange rates, $q$, are defined in the rest
frame of the flow. Seed photons are randomly generated following
the radial distribution of $q_{\rm syn+br}(r)$ and are assumed
to be isotropic in the local rest frame. The flow is assumed to
be uniform tangentially, and thus photons are generated uniformly
in $\theta$. Then, we follow the photon trajectory in
the curved space-time through subsequent scatterings until the
photon either crosses the event horizon or escapes the flow. In
the latter case, it then contributes to the observed
angle-dependent spectrum.

Note that the optical depth in a relativistic flow around a
black hole is affected by both special and general relativistic
effects (see NZ06 for a detailed discussion). The following form
is valid for electron scattering in the Schwarzschild metric, in
a spiral rotating flow with $v^{\theta}=0$,
\begin{eqnarray}
\lefteqn{{\rm d} \tau  = n_{\rm e}
\sigma (E, T_{\rm e}) \gamma   } \\
\lefteqn{\quad \times \left\{ 1 \pm \beta^r \left[ 1 - {\lambda^2 \over r^2} \left(1-{2\over r}\right) \right]^{1\over 2}\! - \beta^{\phi} {\lambda_\phi \over r \sin \theta} \left(1-{2\over r}\right)^{1\over 2} \right\}
\left(1 - {2\over r}\right)^{-{1\over 2}} \!{\rm d} \zeta,\nonumber}
\end{eqnarray}
where $\zeta$ is the affine parameter. The upper and lower sign above corresponds to ingoing (i.e. with ${\rm d} r/{\rm d} \zeta < 0$)
and outgoing photons, respectively (note that $\beta^r <0$).

\begin{figure}
\centerline{\includegraphics[width=7.5cm]{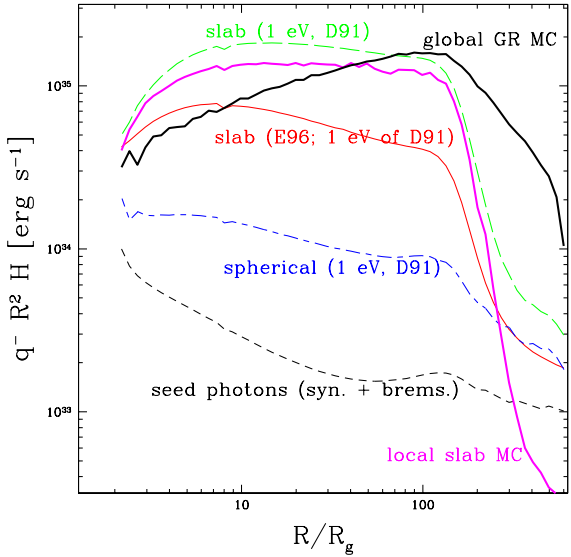}}
\caption{The cooling rates times volume, $q^- R^2 H$, for models
based on the dynamical properties of the initial solution with
$\dot M_0=0.5 \medd$. The black dashed curve gives the rate due
to the synchrotron and bremsstrahlung emission. The thick
magenta and black solid curves correspond to the local slab
model and global GR Compton scattering model, respectively, both
of them obtained using the MC method. The thin red solid
curve gives the cooling rate using the local slab model of E96
but with the optical depth from D91 for their 1-eV seed photon
case. The blue short-long dashes and green dashes correspond to
the local approximation with the model of D91 for a sphere and
slab, respectively, with the coefficients tabulated for 1-eV
seed photons. See Section \ref{compmodels} for details.}
\label{fig:compmodels}
\end{figure}

The simulation of Compton scattering is performed in the local
rest frame using the MC method of G\'orecki \&
Wilczewski (1984; see also Pozdnyakov et al.\ 1983). The
electron temperature, the density, and the velocity vector are
found in each iteration from the hydrodynamical solution by
interpolating between values calculated at 57 logarithmically
spaced radii, $r_i$ (such a grid is sufficient given a smooth
distribution of the flow parameters).

In order to allow us to compare our results with some previous
models, which usually neglect general and special relativistic
effects and use local approximations for Comptonization, we perform
also calculations for local models. We define a set of 57 uniform,
independent spheres or infinite slabs with the $n_{\rm e}$, $T_{\rm
e}$, and the seed spectrum (synchrotron and bremsstrahlung) given
for each  $r_i$. The sphere radius or the slab half-thickness are
given by the corresponding scale height of the flow, $H_i$. Then,
the Thomson optical depth corresponding to the sphere radius or the
slab half depth is $\tau_{{\rm T},i} = \sigma_{\rm T} n_{\rm e}(r_i) H_i$. We then use the MC Comptonization method in flat space for each slab or sphere. The
{\it local} Compton cooling rate can then be determined for each
slab or sphere, while the total luminosity is the sum of the
radiation from all the 57 slabs or spheres.

\section{Comparison of Comptonization Models of Hot Flows}
\label{compmodels}

We first consider some Comptonization models used in other works and
compare them with our results. A quantity often used to quantify the
effect of thermal Compton up-scattering is the amplification factor,
$\eta$, by which the power in initial seed photons is amplified
through Compton scattering. In our formalism,
\begin{equation}
\eta(r)= {q_{\rm syn+br}(r) + q_{\rm comp}(r)\over q_{\rm
syn+br}(r)}. \label{eq:eta}
\end{equation}
In local models, it is assumed that only photons produced at $r$ are Comptonized at this radius. The amplification factor of the Comptonization depends on the local electron temperature, $T_{\rm e}$, the Thomson optical depth, $\tau_{\rm T}$,  and on the initial photon energy, $E_{\rm i}$. Here, we use the formalisms of Esin et al.\ (1996, hereafter E96) and of Dermer, Liang \& Canfield (1991, hereafter D91), averaging their $E_{\rm i}$-dependent $\eta$ over our
seed spectra to get the corresponding Compton cooling rate. We
note whereas E96 give an approximate formula that is supposed to
be valid for any $E_{\rm i}$, equations (1), (20), and table 1
of D91, which we use here, contain coefficients tabulated at
$E_{\rm i}=1$ eV and 1 keV only. Thus, their formula is
accurate only for those values of $E_{\rm i}$ in
spite of the appearance of $E_{\rm i}$ as an argument of their
equation (1).

Fig.\ \ref{fig:compmodels} shows the cooling rates for the dynamical solution given by the initial iteration in Section \ref{results} at $\dot M_0=0.5
\medd$. In this initial iteration only, local slab Comptonization is applied (in order to get the first iteration of the dynamical solution), and the Comptonization cooling (shown as thin red solid curve) is
treated using equation (A10) of E96 and equation (20) of D91,
with coefficients taken from the $E_{\rm i} = 1$ eV slab case. We use here the 1 eV case since most of our synchrotron seed photons are close to this energy. (The results using the original formulae of E96 are very similar to the thin solid curve and thus not shown here.) In addition to Comptonization cooling rates, we also show the cooling rate due to the intrinsic seed emission via synchrotron and bremsstrahlung processes. Then, the ratio of Comptonization cooling rate to the seed one equals to $\eta - 1$. We also show here the
analytical models of D91 for sphere and slab, and compare them to our local NR results and the global GR MC results.

The thick black solid curve shows global GR MC results based on the initial dynamical iteration. We can see that the worst model compared to these accurate results is the spherical one (blue short-long dashes). This is because a sphere is a very poor approximation to the geometry of an accretion flow, more similar to a slab. A key factor affecting Comptonization in hot accretion flows is that the optical depth in the radial direction is much larger than that in the vertical direction (which is $\simeq\! 0.7$ from horizon to the outer boundary for the initial solution). Only models with a slab geometry can account for this property. The assumption of a spherical geometry, which ignores the large difference in the radial and vertical optical depths, leads to significant underestimation of the probability of scattering, and thus under-prediction of both the magnitude of the Compton cooling rate by about an order of magnitude (see Fig.\ \ref{fig:compmodels}), and the hardness of Comptonization spectrum, see Section \ref{results} below.

As we see in Fig.\ \ref{fig:compmodels}, although the local slab model provides much better approximation to the radial profile of the cooling rate than the spherical model, deviations with respect to the global GR calculations remain highly significant. The local slab approximation, in either version of E96 or D91, has several deficiencies, which we discuss below.

First, the seed photons in the formalism of D91 are assumed to
be at either 1 eV or 1 keV. This is obviously not the case in
general. Especially when most of the power in seed photons is at high energies, as it is the case for bremsstrahlung, that
formula becomes highly inaccurate. Since typical electron
temperatures in hot accretion flows are mildly relativistic,
scattering is affected by Klein-Nishina effects. Those effects are also not taken into account by E96. Thus the formulae for the amplification factors from those works usually lead to an overestimation of the Compton cooling rate when bremsstrahlung radiation is the main seed photon source. This can be seen in Fig.\ \ref{fig:compmodels} at $R \ga 200 R_{\rm g}$, where the local analytical results are way above the local slab MC results.

Second, even more importantly, Comptonization is not local. As
shown in Fig.\ \ref{fig:compmodels}, Comptonization can increase the cooling rate by up to two orders of magnitude, i.e., the process is very important. In particular, we see that the cooling rate for global Comptonization is much less centrally concentrated than that for
the local treatments. The seed photon production rate itself is
centrally concentrated; however, those photons travel to outer regions,
increasing the cooling rate there. The local slab model neglects this effect.

Third, strong gravity leads to capture of photons generated in
the innermost region of the accretion flow, and redshifts escaping photons. This will be further discussed in Section \ref{results} below.

Fourth, Comptonization on the flow bulk motion may, in
principle, be important in addition to Comptonization by
electron random motion. Here, we have both rotation and inflow.
The rotation remains slightly sub-Keplerian throughout most of
the flow, and the radial velocity, $v \sim -\alpha
\Omega_{\rm K} R$ (e.g., Narayan \& Yi 1994) is significantly
smaller than the free-fall velocity. Then, both components of
velocity become relativistic only in the innermost
part, within a few $R_{\rm g}$. The bulk plasma motion can
affect the spectrum of the observed radiation through the
following effects. (i) Collimation of the radiation along the
direction of plasma motion, which in a converging flow around a
black hole results in the increase of the number of photons
captured by the black hole. On the other hand, photons emitted
outward have an increased scattering probability due to the
inflow. This increases photon trapping in optically-thick flows.
Even in optically thin flows, photons after those scatterings
are collimated inward, and mostly captured. (ii) The same effect
changes the photon energy after scattering, and the change now
is due to both the bulk and thermal motions of the electrons. A
condition for the bulk motion to be an important source of
energy can be written as $(\gamma-1) m_{\rm e} c^2/(3 k T_{\rm e})
\ga 1$ (e.g., Blandford \& Payne 1981), which is virtually never
satisfied in advection-dominated accretion flows (see
Jaroszy\'nski 2001). In our simulations, we have compared the
power transferred to photons in the plasma rest frame (i.e.,
from thermal motion only) and in the local static frame (with a
contribution of bulk motion), and have found that the latter is
larger by only $\sim\! 10$ per cent greater around $10 R_{\rm
g}$. Only at $\simeq 2.3 R_{\rm g}$ the flow velocity becomes
highly relativistic (and thus motion contributes appreciably),
but the contribution of this region to the total radiated power
is tiny, as seen in Fig.\ \ref{fig:compmodels}. At $\ga 10
R_{\rm g}$, the effect of bulk motion is completely negligible.
Overall, we conclude that the bulk motion Comptonization has
only a minor effect for hot accretion flows.

\section{Effect of outflows and viscous heating of electrons on ADAF solutions}
\label{adaf}

\begin{figure*}
\centerline{\includegraphics[width=6.3cm]{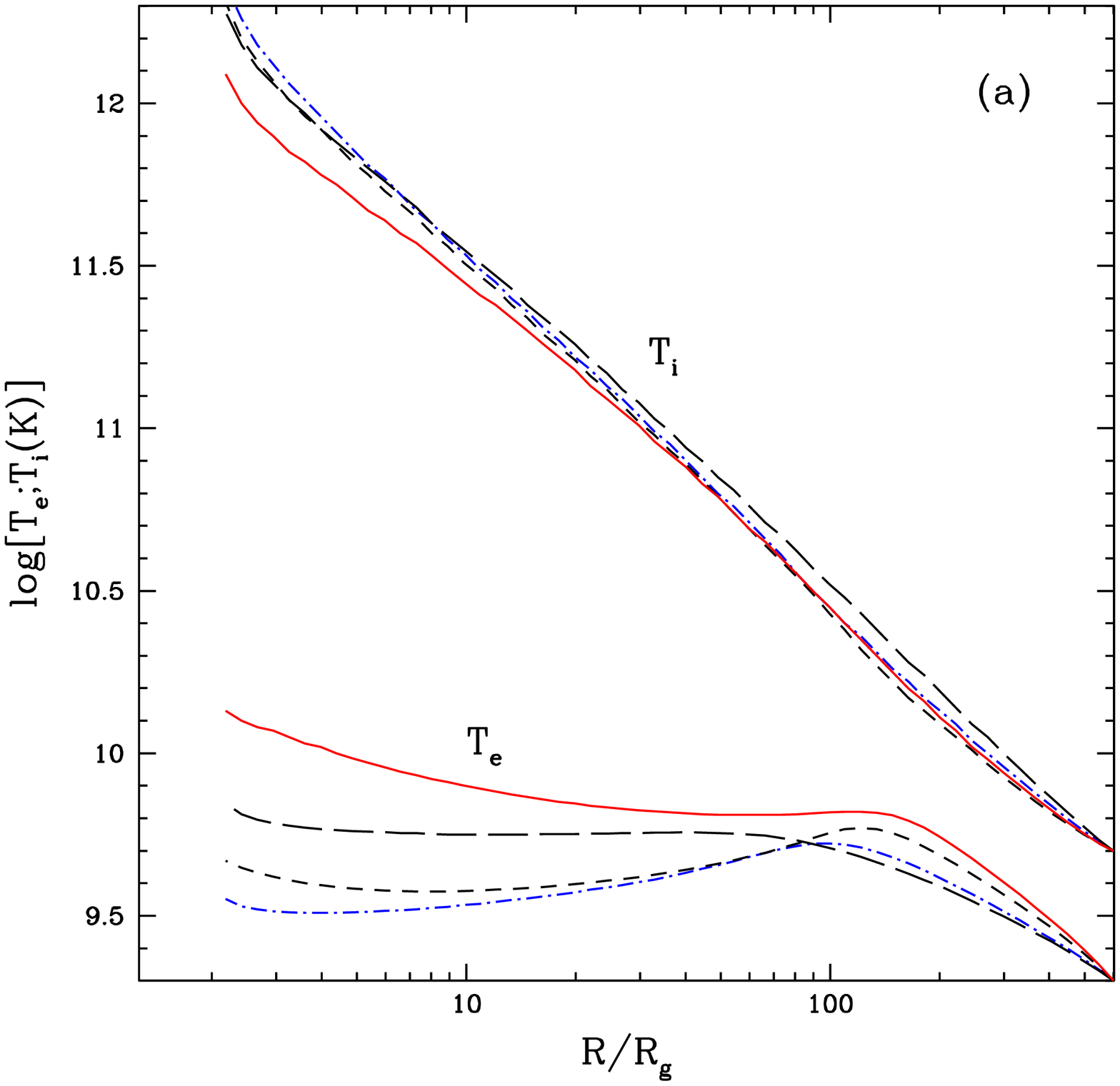}\hspace{0.5cm}
\includegraphics[width=6.3cm]{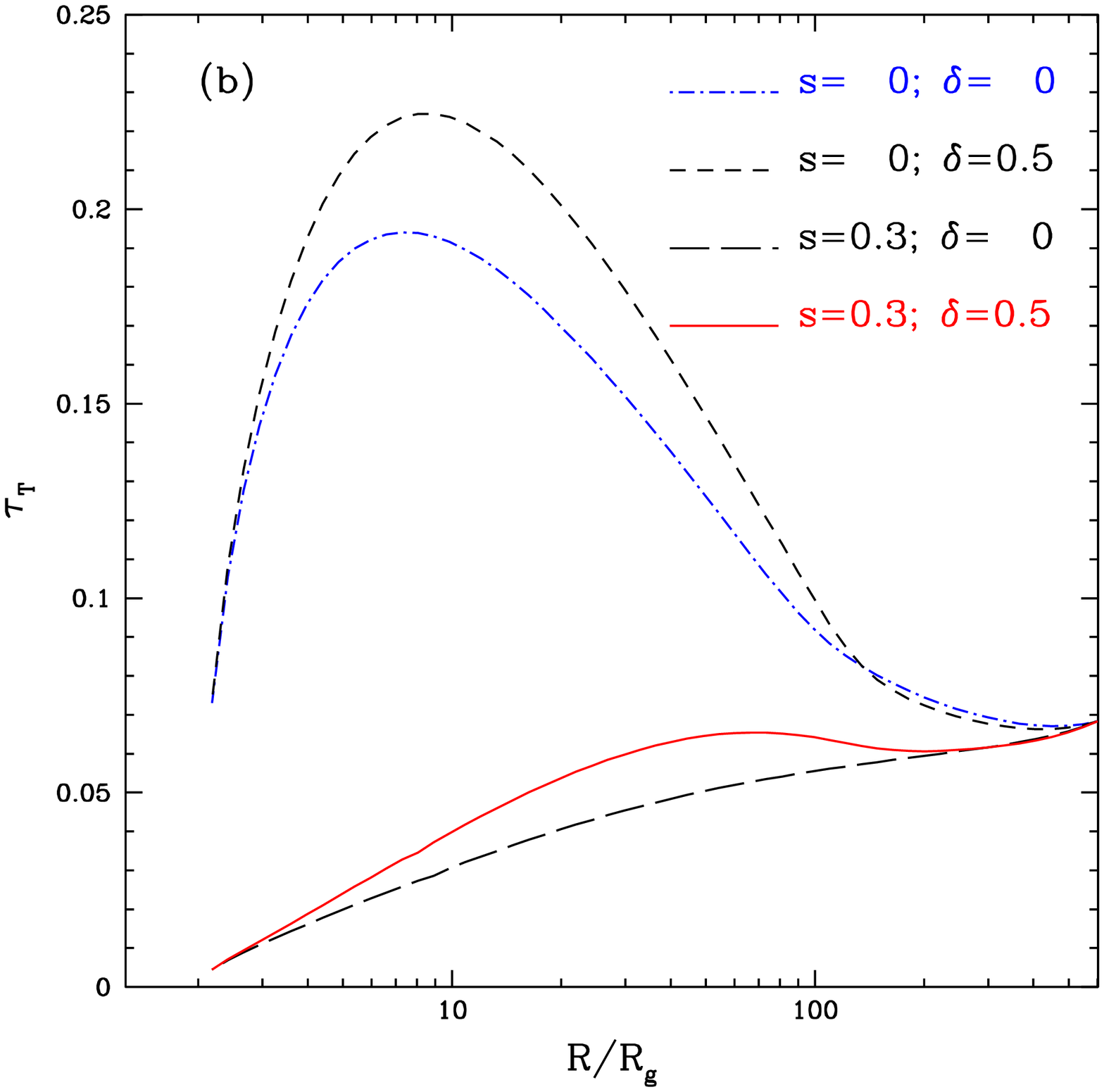}}
\caption{The effect of an outflow (measured by $s$) and viscous heating of electrons (at the fractional rate of $\delta$) on the ADAF solutions. The dot-dashes, short dashes, long dashes and solid curves show the original ADAF model ($s=\delta=0$), and models with strong viscous electron heating and no outflow ($s=0$, $\delta=0.5$), with no viscous electron heating but strong outflow ($s=0.3$, $\delta=0$), with both phenomena included ($s=0.3$, $\delta=0.5$). In all models, Comptonization is treated analytically (D91; E96), and $\dot M_0 = 0.5 \medd$. (a) The radial profiles of the electron (lower curves) and ion (upper curves) temperature. (b) The radial profile of the vertical optical depth.}
\label{fig:outflow}
\end{figure*}

Taking into account of disc outflows (e.g., Stone, Pringle \& Begelman 1999; Blandford \& Begelman 1999) and of viscous heating of electrons (e.g., Bisnovatyi-Kogan \& Lovelace 1997; Quataert \& Gruzinov 1999; Sharma et al.\ 2007) represent two progresses of the hot accretion theory since its initial formulation. Including of either of these possible effects influences accretion solutions in a major way. Since in this work we investigate how global Comptonization affects those solutions, for completeness and to enable comparison among all of these effects, we also show how outflows and viscous electron heating modify the accretion flow structure.

Fig.\ \ref{fig:outflow} shows a comparison between the ADAF model without an outflow, $s=0$, and with no viscous electron heating, $\delta = 0$, and models with either $s$ or $\delta$ or both being $> 0$. These solutions are obtained with local Comptonization using the analytic approximation of E96 and D91 used by us (see Section \ref{compmodels}). We see first that adding viscous electron heating has a relatively minor effect. It somewhat increases $\tau_{\rm T}$ and $T_{\rm e}$ and slightly decreases $T_{\rm i}$. On the other hand, including outflows leads to a dramatic decrease of $\tau_{\rm T}$ and relatively strong increase and decrease of $T_{\rm e}$ and $T_{\rm i}$, respectively, with the effects being much stronger in inner flow regions. Then, setting both $s=0.3$, $\delta=0.5$ results in a superposition of the individual effects.

These effects can be understood as follows. Increasing $\delta$ increases and decreases the power supply to electrons and ions, respectively. An increase of $s$ obviously strongly reduces the density and $\tau_{\rm T}$. This reduces the radiative cooling (roughly $\propto \rho^2$) faster than the viscous heating ($\propto \rho$), which results in an increase of $T_{\rm e}$. The corresponding decrease of $T_{\rm i}$ is due to the reduced compression work caused by the shallower density profile.

\section{Results}
\label{results}

\begin{figure*}
\centerline{\includegraphics[width=6.3cm]{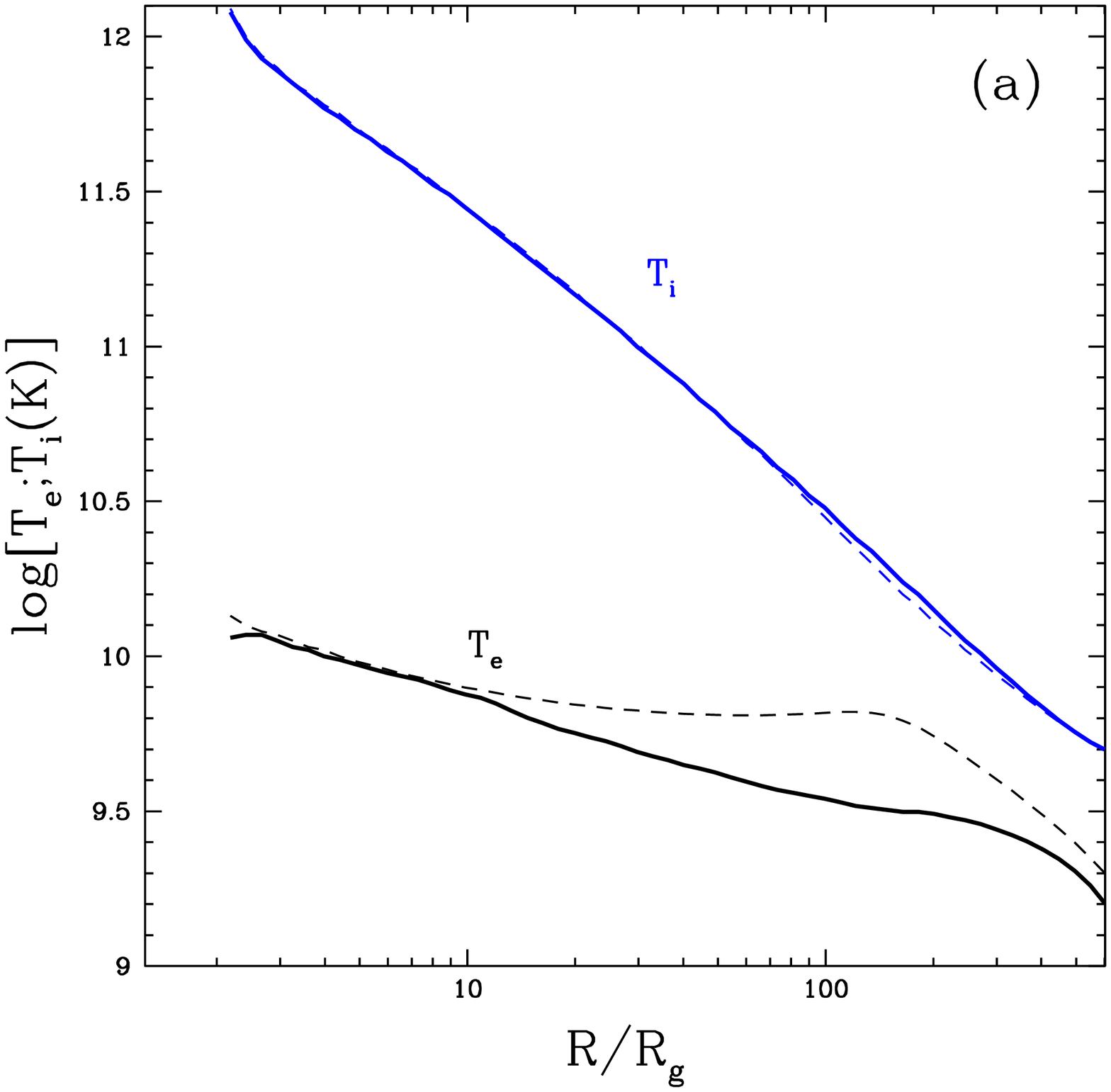}\hspace{0.5cm}
\includegraphics[width=6.3cm]{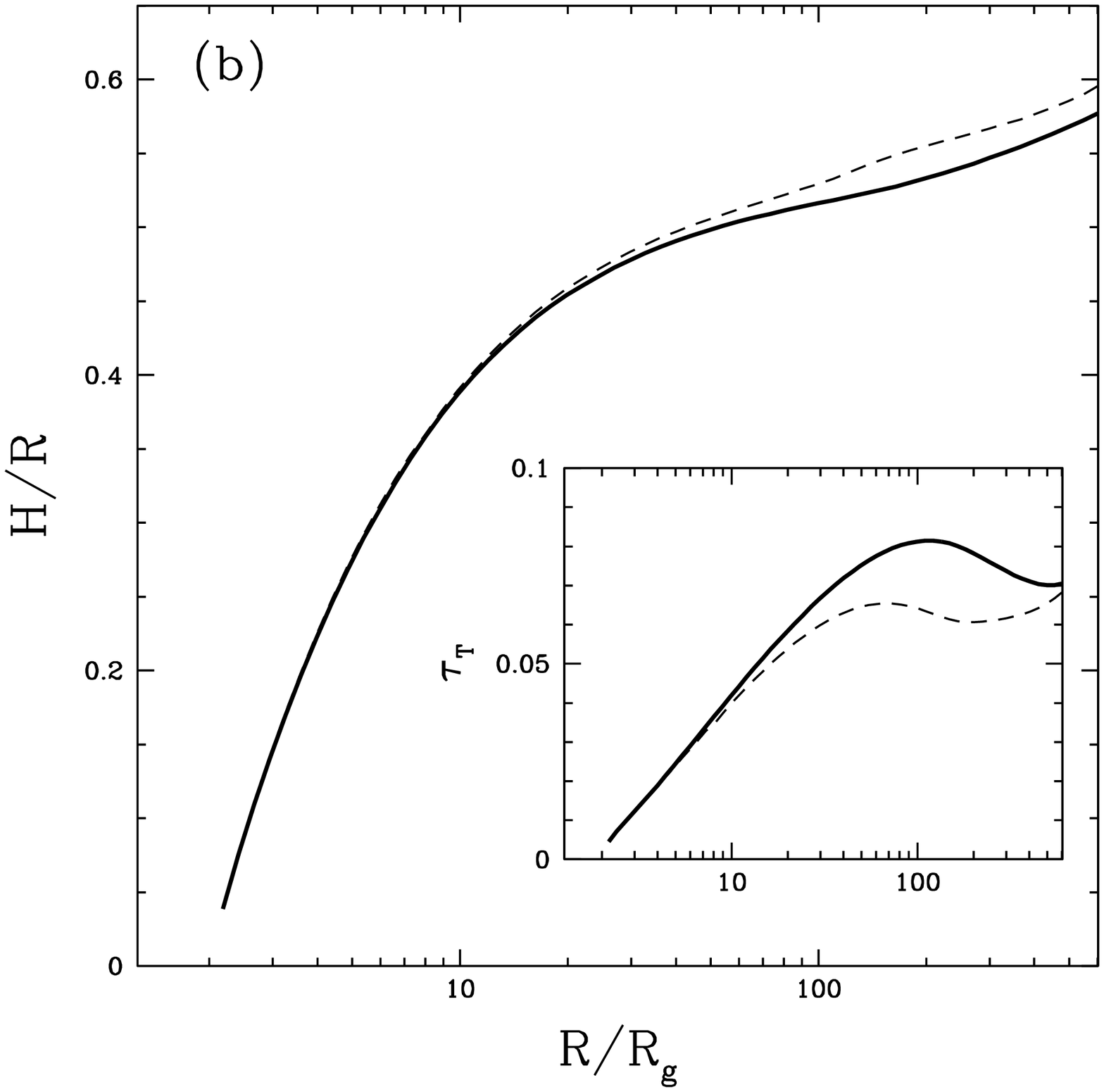}}
\centerline{\includegraphics[width=6.3cm]{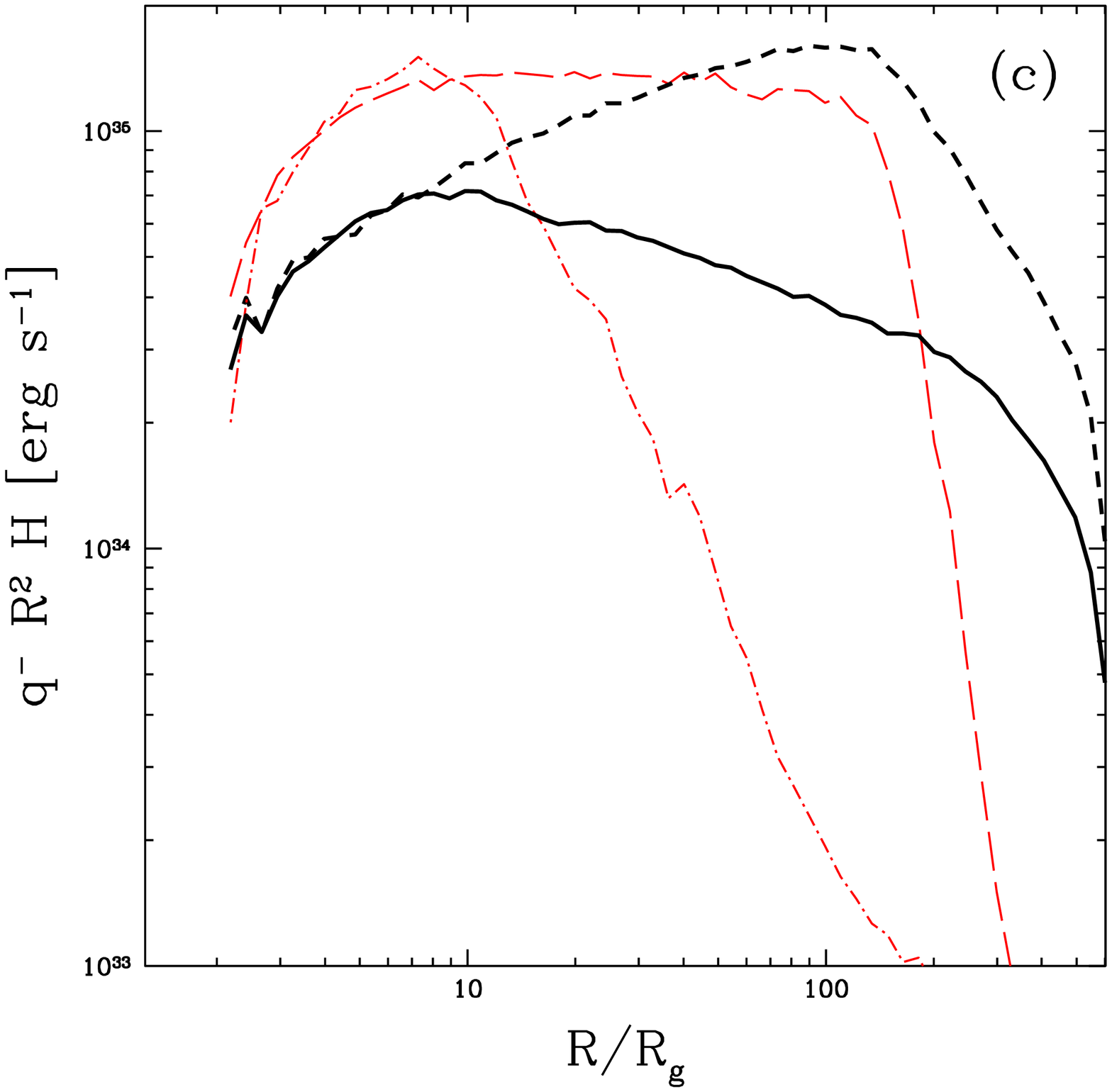}\hspace{0.5cm}
\includegraphics[width=6.3cm]{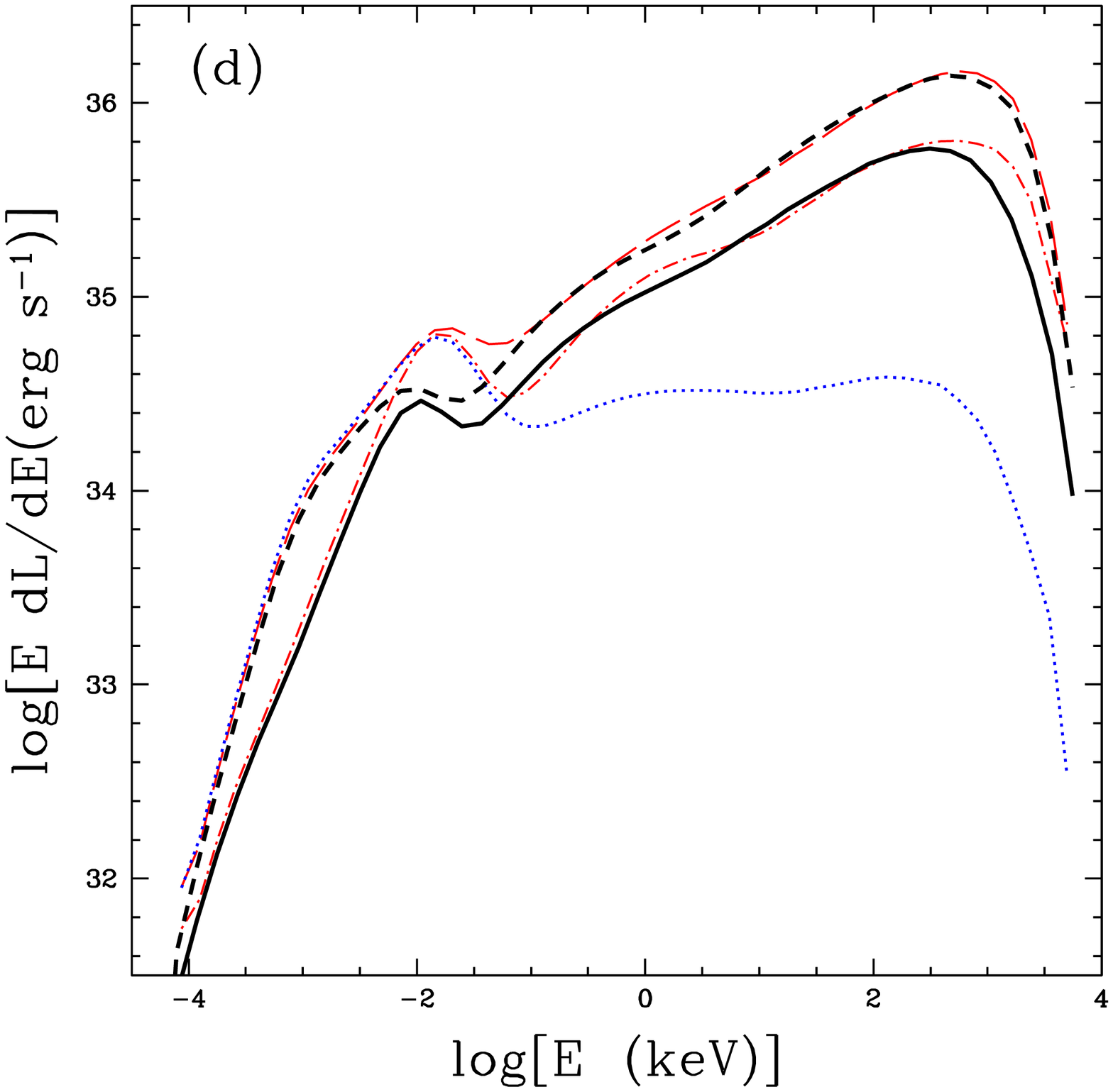}}
\caption{Comparison between the initial (short dashes) and final self-consistent (solid curves) MC solutions with $\dot M_0 = 0.5 \medd$. (a) The radial profiles of the electron (lower curves) and ion (upper curves) temperatures. (b) The ratio of the scale height to the radius, and the vertical optical depth shown in the inset. (c) The radiative cooling rate. (d) The spectrum. In (c--d), the (red) long dashed and dot-dashed curves correspond to local-slab MC simulations of the initial and final solutions, respectively. The blue dotted curve in (d) shows the spectrum from the local-sphere MC simulations for the dynamical structure of the initial solution. [Note that the short and long dashes in (c) are, respectively, the same as the thick solid black and magenta curves in Fig.\ \ref{fig:compmodels}.]} \label{fig:05edd}
\end{figure*}

Fig.\ \ref{fig:05edd} presents comparison between the initial
(short-dashed curves) and the final self-consistent (solid curves) global GR
MC solutions at $\dot M_0=0.5 \medd$. Fig.\ \ref{fig:05edd}(a) shows the
electron and ion, $T_{\rm i}$, temperatures. We see that compared to the initial
solution, $T_{\rm e}$ decreases significantly at $\ga 10R_{\rm
g}$, up to a factor of $\sim\! 2$ at (100--200)$R_{\rm g}$.
This is an effect of the increased cooling in that region, as we
see in Fig.\ \ref{fig:05edd}(c). On the other hand, there is almost no change in $T_{\rm i}$.

The height-to-radius ratio is shown in Fig.\ \ref{fig:05edd}(b).
We find that the height of the flow is nearly unchanged at $\la
20 R_{\rm g}$, and slightly decreases at larger $R$, which is
due to a decrease of $T_{\rm i}$ (and $T_{\rm e}$) there.
We also find only relatively small changes in $v$ and $n_{\rm e}$. The vertical Thomson optical depth shown in the inset of this panel increases slightly, while it remains nearly unchanged at $\la 10 R_{\rm g}$. The horizontal optical depth of the final solution is 0.75.

Fig.\ \ref{fig:05edd}(c) compares the cooling rates for the
global GR model and also for the local slab model. The final
solutions differ significantly from the initial ones. Also, the
global MC solutions differ very strongly from those for
the local slab model. The final self-consistent profile is
relatively flat, namely, drops by only a factor of a few from
the maximum at $\sim\! 10 R_{\rm g}$ to $200 R_{\rm g}$. The flat
profile is partly due to the global Compton scattering. Also,
there is a very strong outflow in our model, see equation
(\ref{eq:outflow}). This strongly reduces both $\dot M$ and the
cooling rate towards the black hole.

Fig.\ \ref{fig:05edd}(d) shows the spectra of the initial and
final solutions from MC simulations with both the local
slab and global GR models. We find that the shape of the
spectrum remains very similar for the initial and final
solutions, but the normalization of the spectra of the final
solutions is about a half of that that of the initial solution.
This is roughly consistent with the result of YXO09. The bolometric luminosity of the final model is $L\simeq 8.4\times 10^{36}$ erg s$^{-1}$, which is $6.7\times 10^{-3}\ledd$. If we define the radiative efficiency as $L/(\dot M_0 c^2)$, it will be $0.013$. However, we stress that most of the potential energy of the flow at $R_{\rm out}$ is lost to the assumed outflow.

For either the initial or final solution, the spectral differences between the global GR and local slab models are relatively small, in particular much smaller than the differences in the cooling rate. This is because each spectrum
is the sum of the radiation produced at all radii, which
averages the local differences. For the final self-consistent
solution, the difference between the global and local models is
significant mostly at the high energy end, $\ga 100$ keV. This
is because this part of the spectrum is produced by
Comptonization in the innermost part of the flow, where the
capture by the black hole and other GR effects
are strong, which reduces the energy of the high-energy cut-off. A
corresponding difference for the initial solutions does not appear
because the high-energy radiation does not originate from the
innermost region, as we can see in Fig.\ \ref{fig:05edd}(c). The differences in the soft X-ray/UV bands
are also caused by the capture and redshift effects. We also see that the local spherical MC model leads to a strong under-prediction of the hardness of the spectrum, see the dotted curve in Fig.\ \ref{fig:05edd}(d).

\begin{figure}
\centerline{\includegraphics[width=6.5cm]{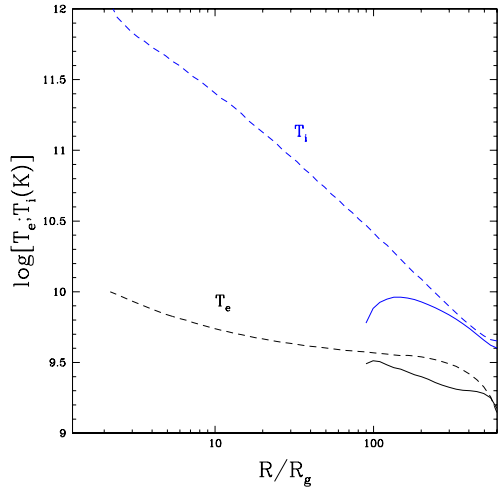}}
\caption{Comparison between the radial profiles of the electron (lower curves) and ion (upper curves) temperatures for MC calculations for the initial (dashed curves) and a solution including global Comptonization (solid curves) for $\dot M_0 = 1.0 \medd$. We see there is no self-consistent solution for $R\la 10^2 R_{\rm g}$ due to the very strong cooling. At lower radii, either the disc collapses or a two-phase cold/hot flow forms.
} \label{fig:10edd}
\end{figure}

When the accretion rate increases (but still $L\ll \ledd$), the radiative cooling will exceed the local viscous heating, and the accretion flow will enter into the LHAF regime (Y01). The accretion flow can still remain hot because
of the heating by the compression work (Y01) and it is the sum
of the viscous dissipation and the compression work that balances
the radiative cooling. Above a certain $\dot M_{\rm crit}$, the
radiative cooling becomes so strong that the flow cannot remain
entirely hot. Such a flow below a certain $R$ (see, e.g., the long
dashes in fig.\ 1 of Y01) will either collapse into a
thin disc or switch into a two-phase flow, with cold clumps embedded
in the hot phase (Y01, Yuan 2003). The value of $\dot M_{\rm crit}$
decreases when the global Compton scattering effect is included.
Here, we find that at $\dot M_0=1.0\medd$, the flow collapses at
$\la 100R_{\rm g}$, as shown in Fig.\ \ref{fig:10edd}. Thus, the
critical accretion rate for our assumed flow with the strong outflow
at $s=0.3$ is given by $0.5<\dot M_{\rm crit}/\medd <1.0$. This $\dot M_{\rm
crit}$ will decrease even further if the cooling due to the photons
from outer thin disc is included. We also find that $\dot M_{\rm crit}$ is higher in the local-Compton model with the cooling rate of E96 and D91 (as used by us above), $\simeq 1.5\medd$ (which solution is not presented here) than in the global-Compton model. This value is then lower than that found by Y01, which is in the range of $\sim\! (3$--$5) \medd$. (Note that Y01 used the definition of $\medd$ including an efficiency factor of 0.1.) The discrepancy
is because both the strong outflow, which suppresses the compression
work (crucial for LHAF), and the strong direct electron heating ($\delta=0.5$)
are included in the present model while they are not in Y01.

\section{Comparison with observations}
\label{comparison}

Data on the spectral hard state of black-hole binaries show the high-energy cut-off around $\sim\! 100$ keV (e.g., Grove et al.\ 1998; Wardzi{\'n}ski et al.\ 2002; Zdziarski \& Gierli\'nski 2004; Done, Gierli\'nski \& Kubota 2007; Joinet, Kalemci \& Senziani 2008). When this cut-off is parameterized by the maximum in the $E {\rm d}F/{\rm d}E$ plots, it is in the range of $\sim\! 50$--200 keV (for spectra with the photon index $<2$, which is the case in the hard state).

We compare these results with our model spectrum, see Fig.\ \ref{fig:05edd}(d). The maximum in the $E {\rm d}L/{\rm d}E$ spectrum of the initial, not self-consistent, solution is at $\simeq 700$ keV if GR effects are neglected and Comptonization is local (the long-dashed curve), which then moves to $\simeq 600$ keV when GR effects are taken into account (the short-dashed curve). The peak energy is at $\simeq 500$ keV for the self-consistent solution when both global Comptonization and GR effects are neglected (the dot-dashed curve). The value of this peak energy is then very similar to the value of the electron temperature, $kT_{\rm e}\sim 700$ keV (Fig.\ \ref{fig:05edd}a) at the radius of the peak dissipation of the corresponding model (Fig.\ \ref{fig:05edd}c). The final dynamic solution has the dissipation rate much more centrally concentrated than the initial one, see Fig.\ \ref{fig:05edd}(c), and then the GR effects reduce the peak energy much more, to $\simeq 300$ keV (the solid curve). Thus, taking into account Comptonization self-consistently, i.e., as the global process in the entire hot disc, and including the GR effects reduces the peak energy of the spectrum by a factor of $\simeq 2/3$. This is because the global Comptonization increases the spectral contribution of outer flow regions while the GR effects reduce the contribution of inner regions. The GR effects here include three parts (see discussion in Section \ref{compmodels}), namely the gravitational and Doppler redshift, and the preferential capture of most energetic photons (enhanced by the kinematic collimation) by the black hole. The last effect is the dominant one.

However, this decrease is still not sufficient to bring the theoretical cut-off down to the observed values. This problem appears to be shared by the hot accretion flow model in general, see, e.g., Yuan \& Zdziarski (2004). As stated above, the cut-off energy before correcting for the GR effects approximately equals the temperature of the flow in the region where most of the radiation is emitted. Emission of this region can then be roughly approximated as that of a one-zone thermal-Comptonization model. Fits of this model to hard-state spectra of black-hole binaries and of Seyfert galaxies show $kT_{\rm e}\simeq 50$--100 keV, see, e.g., the compilation in Yuan \& Zdziarski (2004). Approximately, $kT_{\rm e}$ of the one-zone model equals the energy of the maximum of its $E {\rm d}L/{\rm d}E$ spectrum. The other parameter of this model is the Thomson optical depth, $\tau_{\rm T}$. For a given X-ray spectral index, the Compton parameter, $y=4 T_{\rm e}\tau_{\rm T}$, is approximately constant. Also, $kT_{\rm e}$ is anticorrelated with $\tau_{\rm T}$ for a given local dissipation rate in a hot flow, reflecting the varying power per electron. Fits of the one-zone thermal-Compton model yield $\tau_{\rm T}\sim 1$ (e.g., Yuan \& Zdziarski 2004) whereas the vertical optical depths of the flow are $\tau_{\rm T}\ll 1$, see Fig.\ \ref{fig:05edd}(b), which expresses the same discrepancy as $T_{\rm e}$ of the flow being too high. As stated above, taking into account the GR effects in the global-Compton model improves the agreement significantly (pointing to the inadequacy of Comptonization models neglecting GR), but not sufficiently.

On the face of it, this discrepancy might indicate that the hot flow model is not applicable to the hard state of black hole binaries and Seyfert galaxies. However, we note that the present calculations assume a very strong outflow, equation (\ref{eq:outflow}) with $s=0.3$, motivated by the modelling of the Galactic Centre (Yuan et al.\ 2003). On the other hand, our assumed dimensionless accretion rate is $\gg$ that of the Galactic Centre source, whereas the relative strength of outflows appears to decrease with the increasing accretion rate (Gallo, Fender \& Pooley 2003; Fender, Belloni \& Gallo 2004), as also indicated by a relatively small change of the X-ray bolometric luminosity during hard/soft state transitions (e.g., Zdziarski et al.\ 2004). This is plausible from a theoretical point of view because the Bernoulli parameter becomes smaller with the increasing accretion rates (Yuan, Cui \& Narayan 2005; Bu \& Yuan in preparation). Thus, it is likely that the fractional strength of the outflow, 82 per cent for our assumed $s$, $R_{\rm out}$ and $\dot M_0$ (with 18 per cent of $\dot M_0$ crossing the horizon), may in reality be much weaker. Setting $s=0$ leads to $\tau_{\rm T}$ being a few times higher and $T_{\rm e}$ about a factor of two lower, see Fig.\ \ref{fig:outflow}. Although the calculations in Fig.\ \ref{fig:outflow} are done with the local analytical treatment of Comptonization, and are thus not self-consistent, they correctly predict the direction of the changes of the flow parameter. If the characteristic $T_{\rm e}$ of the self-consistent flow were reduced by the same factor of two with respect the model with $s=0.3$, the peak of the $E {\rm d}L/{\rm d}E$ spectrum would also go down by a similar factor, bringing it to the observed range and resolving the discrepancy with the data.

Our second assumption has been of a strong viscous heating of electrons, $\delta=0.5$. As seen in Fig.\ \ref{fig:outflow}, this has a relatively minor effect on $T_{\rm e}$ and $\tau_{\rm T}$. We also note that the half-depth Thomson optical depth of a slab is the quantity closest to that of the half-depth of the flow, and should preferably be used when comparing accretion flow models with one-zone thermal Comptonization models. For a given X-ray spectral index, the optical depth is somewhat lower for a slab geometry than for spherical one.

We note that there is a number of additional effects that can further reduce $T_{\rm e}$ and increase $\tau_{\rm T}$, possibly allowing the hot flow model to be in agreement with the full range of the observed hard-state spectra. First, an increase of the black-hole spin reduces the radial velocity and increases the density of the flow. The effect is rather strong, as can be seen, e.g., in fig.\ 5 of Gammie \& Popham (1998). Global Comptonization in the Kerr metric will be studied in detail in our forthcoming work. Second, the presence of moderate large-scale toroidal magnetic fields in the accretion flow significantly reduces $T_{\rm e}$, as shown by Bu, Yuan \& Xie (2009). Third, electron cooling will also be significantly enhanced in either a two-phase flow, with cold clouds mixed with the hot flow, or, fourth, at presence of an inner collapsed disc, with both effects happening above some critical accretion rate (Section \ref{results}, see also Y01, Yuan 2003).

Furthermore, the cut-off energy in the hard state of black-hole binaries is observed to decrease with the increasing luminosity (Wardzi{\'n}ski et al.\ 2002; Yamaoka et al.\ 2006; Yuan et al.\ 2007; Miyakawa et al.\ 2008)\footnote{Note that temperatures as low as $\sim\! 20$ keV were claimed in the one-zone model fits in Miyakawa et al.\ (2008). However, they were obtained using the non-relativistic Comptonization model of Sunyaev \& Titarchuk 1980), which is not valid for spectra extending above $\sim\! 50$ keV.}. Also, the hot-flow characteristic temperature goes down with the increasing accretion rate, in agreement with the observations. In our case, the bolometric luminosity is only $0.007 L_{\rm E}$ (Section \ref{results}), and thus our model spectrum should be compared with the hard state at the correspondingly low $L$ (which is somewhat below, e.g., the hard-state range of $L$ of Cyg X-1 of $\simeq 0.01$--$0.02L_{\rm E}$, Zdziarski et al.\ 2002), which spectra are likely to have the cut-off energies $>100$ keV. This would be in agreement with our results, after the correction for the outflow discussed above.

Malzac \& Belmont (2009) have also pointed out that the ion temperatures implied by observations of the hard state are much lower than those typical for ADAF models. However, the ratio of $T_{\rm i}/T_{\rm e}$, calculated by Malzac \& Belmont (2009) directly from the formula for Coulomb energy transfer from ions to electrons, is approximately $\propto \tau_{\rm T}^{-2}$, and the discrepancy pointed out by those authors occurs at $\tau_{\rm T}\sim 1$. If we take instead the low values of $\tau_{\rm T}$ obtained in ADAF models, there is an agreement between their estimate and the hot flow models.

On the other hand, we stress that using equation (\ref{eq:outflow}) with $s=0.3$ does not weaken our conclusions related to the role of global Comptonization. Instead, if the outflow is weaker for a given $\dot M_0$, the radiation generated at small radii and subsequently received at large radii will be stronger, thus the global Comptonization effect will become even more important than for our chosen assumption. This would further reduce the average photon energy of the emitted spectrum.

Another related issue is our determination of the value of the
critical accretion rate, at which an inner part of the hot flow
collapses. We find it corresponds to a relatively low $L\simeq
0.01\ledd$. Luminosities of the hard state of black-hole binaries
are commonly above this $L$ (e.g, Done et al.\ 2007; Zdziarski et
al.\ 2002). Again, our determination is for the assumed strong
outflow, and relaxing this assumption may increase that critical
luminosity. On the other hand, the presence of a collapsed
geometrically thin disc close to the horizon would explain the
finding of relativistically broaden Fe K$\alpha$ line in the hard
state of black-hole binaries (e.g., Miller et al.\ 2006; Miller 2007; but see Done \& Diaz Trigo 2009 for a critical view), whose origin would have otherwise been in conflict with the hot accretion flow model of the hard state.

Finally, we note that alternative models for the hard state have been proposed. A very interesting recent model is non-thermal, in which the power supplied to electrons (with the optical depth of $\tau_{\rm T}\ga 1$) goes into their acceleration into a power-law distribution (Poutanen \& Vurm 2009; Malzac \& Belmont 2009). Then, synchrotron self-absorption and Coulomb interactions efficiently thermalize the electrons provided any blackbody emission is weak. This radiative one-zone model fits the hard-state data, e.g., of Cyg X-1, very well. An important issue here is the location of the non-thermal plasma. It cannot be a corona as the model constrains any disc blackbody emission irradiating the plasma to be very weak. If it is an inflow, it would require an efficient mechanism of electron acceleration (instead of heating) with a large $\delta$, and a low radial velocity to achieve the required large $\tau_{\rm T}$. It appears unclear whether such a model is possible dynamically.

\section{Conclusions}
\label{conclusions}

Hot accretion flows, both ADAF and LHAF, are optically thin in
the radial direction, and thus the global Compton scattering is
important. Its effect on the dynamics of the accretion flow has
been investigated previously by an analytical approach in YXO09.
In this paper we revisit this problem using the more accurate
MC simulations, and focusing on the inner region of the
flow. We confirm that the global Compton scattering effect is
dynamically very important. We obtain the final self-consistent
solution using the iteration approach. Our main results, in
particular showing the differences between the initial and final
iterations, are presented in Fig.\ \ref{fig:05edd}. The global
GR cooling rate of the final self-consistent solution is more
centrally concentrated than that of the initial solution, but
less centrally concentrated than the local cooling rate of the
final solution.

To evaluate the effect of global Compton scattering, we also
compare the global GR Compton scattering model with two local
models, in the sphere and slab geometries. In these
models, only the local scattering is taken into account. For
both the initial and final self-consistent solutions, we find
that none of the local models can adequately reproduce the
global model, although the slab model is significantly better
than the spherical one. This is due to the fact that the radial optical depth of the hot flow is much higher than the vertical one.

Specifically, the local slab model underestimates the cooling
rate in an outer region, $\ga 200 R_{\rm g}$, by up to two
orders of magnitude with respect to the global GR MC
results, while it overestimates the cooling rate by up to factor
of a few in an inner region, $\la 20 R_{\rm g}$. This
differences occur because of the transfer of seed photons
between different radii, mostly from small to large radii. We
also find that the main dynamical effect in our self-consistent treatment
is the global Comptonization, whereas the GR and bulk motion effects
are dynamically less important.

We also compare the spectra of the initial and final solutions in both
the local and global models. We find that the shape of the spectrum of the final
self-consistent solution changes only slightly (mostly in the reduced high-energy cut-off) compared to that of the initial solution, but the normalization decreases by a factor of $\sim 2$. We also find that the spectral differences between global and local slab models are much less significant than those in the radial profile of the cooling rate. This is due to the averaging of the local spectra over all radii.

We have compared our results to the spectra of the hard state of black-hole binaries, see Section \ref{comparison}. We find an overall rough agreement, though our models have the characteristic electron temperatures and the characteristic Thomson optical depths higher and lower, respectively, than those required by the observational data. However, this discrepancy can, most likely, be resolved if the disc outflow in the luminous hard states is much weaker than that assumed by us. Additional factors that would allow the model be in agreement with the full range of the observed spectra are the spin of the black hole, large-scale magnetic fields, and the presence of an additional electron cooling mechanism, e.g., soft photons from cold clouds embedded in the hot flow, or from an inner collapsed thin disc. We stress that global Comptonization is crucial in bringing the hot accretion flow model to agreement with the data. This is due to the self-consistent model having the dissipation much more centrally concentrated than that of local models, which in turn causes the GR effects to reduce the spectral cut-off energy much more for the former than the latter.

\section*{Acknowledgments}

FGX and FY have been supported in part by the Natural Science
Foundation of China (grants 10773024, 10821302, 10825314 and 10833002),
One-Hundred-Talent Program of Chinese Academy of Sciences, and
the National Basic Research Program of China (grant
2009CB824800). AAZ and AN have been supported in part by the Polish
MNiSW grants NN203065933, 362/1/N-INTEGRAL/2008/09/0 and N20301132/1518.

\label{lastpage}

\end{document}